\documentclass[twocolumn,showpacs,preprintnumbers,amsmath,amssymb]{revtex4}

\usepackage{amsmath}
\usepackage{graphicx}
\usepackage{dcolumn}
\usepackage{bm}
\usepackage{CJK}
\usepackage{mathrsfs}
\usepackage[usenames]{color}
\usepackage{ulem}

\begin{document}
\begin{CJK*} {UTF8} {gbsn}

\preprint{1}



\title{$\Omega$-dibaryon production with hadron interaction potential from the lattice QCD in relativistic heavy-ion collisions}



\author{Song Zhang(张松) and Yu-Gang Ma(马余刚)\footnote{Email: mayugang@fudan.edu.cn}}
\affiliation{Key Laboratory of Nuclear Physics and Ion-beam Application~(MOE), Institute of Modern Physics, Fudan University, Shanghai $200433$, China}

\date{\today}

\begin{abstract}
Recently the HAL QCD Collaboration reported the $\Omega-\Omega$ and $N-\Omega$ interaction potentials by the lattice QCD simulations. Based on these results, $N\Omega$ ($^5S_2$) and $\Omega\Omega$ ($^1S_0$) bound states were predicted with the  binding energy about a few MeV. In addition, $N-\Omega$ HBT correlation function was also  measured by the STAR Collaboration as well as the ALICE Collaboration. These results provide dynamical information whether or not $\Omega$-dibaryons exist in the interaction aspects. Another necessary point for the detection of $\Omega$-dibaryons is the experimental environment where the bound state could be produced and survived in the system.
In this context,  there are at least two necessary conditions to constrain the production probability of $\Omega$-dibaryons, i.e. the one is the necessary short-range attractive interaction to form the bound state and the another is the  experimental environment such as heavy-ion collision provides abundant enough strangeness and multiplicity of nucleons. In this Letter the $\Omega-\Omega$ and $\Omega-$nucleon interaction potentials by the lattice QCD simulations were employed to obtain $\Omega\Omega$ ($^1S_0$) and $N\Omega$ ($^5S_2$)  wave functions, and then the productions of $\Omega$-dibaryons were estimated by using of a dynamical coalescence mechanism for the relativistic heavy-ion collisions at $\sqrt{s_{NN}} = $ 200 GeV and 2.76 TeV.
\end{abstract}





\maketitle

\section{Introduction}
Dibaryon is attracting much attention in hadron physics as well as heavy-ion physics communities.  In traditional sector of light quarks, only one stable dibaryon was  experimental  measurable so far,  i.e. deuteron ($d$) as a molecular state of neutron and proton. In sector of strange quark, strangeness dibaryon has been investigated in theory and experiments for a long period since $H-$dibaryon was predicted by Jaffe~\cite{PhysRevLett.38.195}.  $\Omega-$dibaryons, such as  $N\Omega$ ($^5S_2$) and $\Omega\Omega$ ($^1S_0$), were proposed in several theoretical work and considered as the most promising candidates of strangeness dibaryons~\cite{CLEMENT2017195, CHO2017279}. 
Goldman {\it et~al.}~\cite{PhysRevLett.59.627} predicted the strangeness-3 dibaryons by using of two different quark models of hadrons. In the framework of the quark delocalization color screening model and the chiral quark model,  the $N\Omega$ dibaryon was further studied and its  binding energy was estimated from a few MeV to a hundred MeV within different configurations~\cite{PhysRevC.51.3411,PhysRevC.92.065202}. A baryon-baryon interaction model with meson exchanges also calculated the $N-\Omega$ two-body system which was suggested as a quasibound state with the binding energy of 0.1 MeV~\cite{PhysRevC.98.015205}.  The lattice QCD near the physical point (with pion mass $m_{\pi}$ = 146 MeV) suggested that the binding energy of $N\Omega$ dibaryon is 2.46 MeV and 1.54 MeV, respectively, with and without Coulomb attraction~\cite{IRITANI2019284}.


$\Omega\Omega$ dibaryon with strong $\Omega-\Omega$ attraction was predicted by a chiral quark model~\cite{PhysRevC.61.065204,LI2001487,Dai}, while  with a weak repulsion $\Omega-\Omega$ interaction was suggested by other models~\cite{PhysRevC.51.3411,PhysRevD.85.094511}. Based on the possible production channel~\cite{You_Wen_2001,PhysRevC.66.015205}, an extended version of a multi-phase transport (AMPT) model~\cite{PAL2005210} estimated the production probability of $\Omega\Omega$ dibaryon in Au + Au collisions at $\sqrt{s_{NN}}$ = 130 GeV. The HAL QCD method with pion mass close to the physical point ($m_{\pi}$ = 146 MeV) presented $\Omega\Omega$ binding energy of 1.6 MeV or 0.7 MeV without/with the Coulomb repulsion~\cite{PhysRevLett.120.212001}, respectively.


The momentum correlation function of hadron pairs can reflect the hadron-hadron interaction~\cite{STAR-diLambda,STAR-dipbar,Chen} and provide the information whether the pair can form a bound state or not. Based on the lattice QCD simulations of $N-\Omega$ interaction~\cite{NPA201489}, the momentum correlation function of $N-\Omega$ was calculated in Ref.~\cite{PhysRevC.94.031901} for providing information on possible $N\Omega$ dibaryon bound state. The STAR collaboration conducted this measurement~\cite{STAR-pO-2019490} in Au+Au collisions at $\sqrt{s_{NN}}$ = 200 GeV and the results favored the p-$\Omega$ bound state with the binding energy of 27 MeV. By using of the interaction potential  from the recent lattice QCD calculations at near physical quark masses~\cite{PhysRevLett.120.212001,IRITANI2019284}, the $p-\Omega$ and $\Omega-\Omega$ momentum correlation functions were updated~\cite{PhysRevC.101.015201}. Recently, the ALICE collaboration reported the measurement of $p-\Omega^{-}$ and $p-\Xi^{-}$ correlation functions~\cite{2020arXiv200511495A} in $pp$ collision at $\sqrt{s}$ = 13 TeV, and the result of $p-\Xi^{-}$ was in agreement with the predicted correlation function by the HAL QCD results ~\cite{PhysRevC.101.015201}.  Furthermore, $p-\Omega^{-}$ correlation function should be investigated for  nucleus-nucleus collisions in theoretical and experimental aspects.

In this Letter, we reported the production of $N\Omega$ ($^5S_2$) and $\Omega\Omega$ ($^1S_0$) dibaryons calculated by a dynamical coalescence model with the consideration of $N-\Omega$ and $\Omega-\Omega$ interaction potentials from the HAL QCD results ~\cite{IRITANI2019284,PhysRevLett.120.212001} for  Au+Au collisions at $\sqrt{s_{NN}}$ = 200 GeV and Pb+Pb collisions at $\sqrt{s_{NN}}$ = 2.76 TeV. It is found that the production probabilities of $N\Omega$ ($^5S_2$)  and of $\Omega\Omega$ ($^1S_0$)  are about $\sim10^{-3}$ and $\sim10^{-6}$, respectively.

\section{A brief introduction to algorithm}

\subsection{Blast-wave model calculation}

Dynamical coalescence model is able to describe hadron and light nuclei productions in heavy-ion collisions ~\cite{PhysRevLett.84.4305,CHEN2003809,ZHANG2010224,SUN2015272,PhysRevC.95.044905}, in which the constituent interaction was reflected in the relative wave function. For two-body clustered object, the multiplicity of the object can be obtained by~\cite{PhysRevLett.84.4305,CHEN2003809,ZHANG2010224,SUN2015272,PhysRevC.95.044905},
\begin{eqnarray}
\begin{split}
N_{2b} &= g_{2} \int \left(d^4x_1 S_1(x_1,p_1)\frac{d^3p_1}{E_1}\right)\times\\
&\left(d^4x_2 S_2(x_2,p_2)\frac{d^3p_2}{E_2}\right)\times \rho^{W}_2(x_1,x_2; p_1,p_2),
\end{split}
\label{coal_ana_def}
\end{eqnarray} 
where $\rho^{W}_2(x_1,x_2; p_1,p_2)$ is the Wigner density function which gives the coalescence probability, $g_2 = (2S+1)/(2s_1+1)(2s_2+1))/ N_I$ is the coalescence statistical factor~\cite{POLLERI1999452}, $S$ is spin of the clustered object and $s_i$ ($i$=1,2) the spin of the two constituents, $N_I$ counts for the isospin states and $N_I(d)$ = 2 for deuteron ($d$), 1 for others in this work, and then $g_2$ is  3/8 for $d$, 5/8 for $N\Omega$ and 1/16 for $\Omega\Omega$, respectively. Note that the isospin contribution to the coalescence statistical factor is still an open question, i.e. whether it should be included or   not in the coalescence calculation. Some works included this contribution~\cite{CHEN2003809,POLLERI1999452,PhysRevLett.84.4305,ZHANG2010224} but some did not~\cite{PhysRevC.98.054905-isospinNO,PhysRevC.92.064911-isospinNO,SUN2019132-isospinNO,2020arXiv200500182S}. $S(x,p)$ is the phase space distribution of the constituents at coordinate $x$=$(\vec{r}, t)$ and momentum $p$=$(\vec{p}, E)$. The two-body's position and momentum were taken at equal time in their rest frame. Note that the phase space distribution of neutron ($n$) was assumed the same as that of proton ($p$) if a certain cluster contains a neutron.

The phase-space distribution can be expressed by a Blast-wave model~\cite{PhysRevC.70.044907,SUN2015272,PhysRevC.95.044905,PhysRevC.89.034918},
\begin{eqnarray}
\begin{split}
S(x,p)d^4x &= m_T\cosh(\eta_s-y_p) f(x,p) J(\tau)\times\\
&\tau d\tau d\eta_s rdrd\varphi_s,
\end{split}
\label{phase_space_def}
\end{eqnarray} 
where $y_p$ and $m_T$ are respectively the rapidity and transverse mass of the hadron, $r$, $\tau$, $\eta_s$, and  $\varphi_s$ are the polar coordinates, proper time, pseudorapidity and azimuthal angle in coordinate space, respectively. 
A Gaussian distribution for the freeze-out proper time is given by, $J(\tau)=\frac{1}{\Delta\tau\sqrt{2\pi}}\exp[-\frac{(\tau-\tau_0)^2}{2(\Delta\tau)^2}]$, where $\tau_0$ and $\Delta\tau$ are the mean value and the dispersion of the $\tau$ distribution. The statistical distribution function~\cite{PhysRevD.10.186} $f(x,p)$ is defined by $f(x,p)$ = $\frac{2s+1}{(2\pi)^3}\left[\exp\left(p^{\mu}u_{\mu}/T_{kin}\right)\pm1\right]^{-1}$, where $s$ is the spin of the particle, $u_{\mu}$  is the four-velocity of a fluid element in the fireball of the emission source, and $T_{kin}$ is the kinetic freeze-out temperature. The energy in the local rest frame of the fluid can be written as, $p^{\mu}u_{\mu}$ = $m_T\cosh\rho\cos(\eta_s-y_p)-p_T\sinh\rho\cos(\varphi_p-\varphi_s)$, where $\varphi_p$ is azimuthal angle in momentum space, $\rho$ is the transverse flow rapidity distribution of the fluid element in the fireball with a transverse radius $R_0$, defined as $\rho$ = $\rho_0\frac{r}{R_0}$ without considering the anisotropic part~\cite{PhysRevC.70.044907,SUN2015272,PhysRevC.89.034918}. Once  the parameters of ($\tau_0$, $\Delta\tau$, $\rho_0$, $R_0$, and $T_{kin}$) are fixed, one can obtain the transverse momentum distribution of a hadron by,
\begin{eqnarray}
\frac{dN}{2\pi p_Tdp_Tdy_p} &=& \int S(x,p)d^4x.
\label{pT_spectra_def}
\end{eqnarray} 

\subsection{AMPT model calculation}

AMPT model can provide the phase-space data of  constituent particles with which the dynamical coalescence model for two-body clustered objects can be performed  by the following Eq.~\cite{CHEN2003809,ZHANG2010224},
\begin{eqnarray}
N_{2b} = g_{2} \int \left<\sum \rho^{W}_2(x_1,x_2; p_1,p_2)\right>\times d\vec{r}_1d\vec{r}_2 d\vec{p}_1d\vec{p}_2,
\label{coal_tranport_def}
\end{eqnarray} 
where $\left<...\right>$ denotes event averaging and the sum runs over all possible combinations of the two bodies. Coordinate $x$ = $(\vec{r}, t)$ and momentum $p$ = $(\vec{p}, E)$ for each of the two bodies can be obtained from the AMPT transport model.

In the present simulation of the AMPT model~\cite{AMPT2005}, the  version $2.26t7b$ was employed to provide the phase-space information of neutrons, protons and $\Omega$'s. AMPT simulates the relativistic heavy-ion collisions dynamically in a framework of multi-phases, namely partonic phase and hadronic phase,  in which the initial phase is given by the Heavy Ion Jet Interaction Generator (HIJING) model~\cite{HIJING-1,HIJING-2}, the melted partons from the HIJING interact with each other by the Zhang's Parton Cascade (ZPC) model~\cite{ZPCModel}, and  the interacting-ceased partons will finally convert to hadrons by a simple quark coalescence model or the Lund string fragmentation model, then hadrons experience rescattering  by a relativistic transport model (ART) ~\cite{ARTModel}. The AMPT model can well describe different physics for relativistic heavy-ion collisions at the RHIC~\cite{AMPT2005} as well as the LHC~\cite{AMPTGLM2016} energies, eg. for hadron HBT correlations~\cite{AMPTHBT}, di-hadron azimuthal correlations~\cite{AMPTDiH,WangHai}, collective flows~\cite{STARFlowAMPT,AMPTFlowLHC}, strangeness productions~\cite{NSTJinS,SciChinaJinS} as well as chiral magnetic effects and so on \cite{Zhao,Huang,Wang,XuZW}.

Both the Blast-wave model and the AMPT model can calculate dibaryons by coupling with coalescence model by Eqs.~(\ref{coal_ana_def}) and~(\ref{coal_tranport_def}), respectively. The Blast-wave model is controlled by some parameters, namely $T_{kin}$, $\rho_0$, $\tau_0$, $\Delta\tau$, and $R_0$, which presents the bulk properties of the collision system at kinetic freeze-out stage. As a transport model, the AMPT model considers the microscopic transport process, such as partonic interaction and hadronic rescattering, it can consequently investigate the dynamical process of the heavy-ion collisions except for providing phase-space of particles as an event generator in the present work.

\subsection{Wigner density function}

The Wigner density function of the objects ($\rho^{W}_2$) in Eq.~(\ref{coal_ana_def}) and Eq.~(\ref{coal_tranport_def}) can be obtained from,
\begin{eqnarray}
\begin{split}
\rho^W_2(\vec{r},\vec{q}) = \int \phi\left(\vec{r}+\frac{\vec{R}}{2}\right)\phi^{*}\left(\vec{r}-\frac{\vec{R}}{2}\right)\times\\
\exp\left(-i\vec{q}\cdot\vec{R}\right)d\vec{R},
\end{split}
\label{wigner_density_def}
\end{eqnarray} 
where $\vec{q}$ = $(m_2\vec{p}_1-m_1\vec{p}_2)/(m_1+m_2)$ and $\vec{r}$ = $(\vec{r}_1-\vec{r}_2)$ are the relative momentum and relative coordinate, respectively, $\phi(\vec{r})$ is the relative wave function of the two constituents. In our previous work for three constituents and other calculations for two bodies, the relative wave function was always taken to be a spherical harmonic oscillator. In this work, the wave functions and binding energies $E_{B}$ for $d$ ($^3S_1$), $N\Omega$ and $\Omega\Omega$ at assumed bound states of $^5S_2$ and $^1S_0$ will be obtained by solving the radial Schr\"odinger equation with the potentials for $n-p$~\cite{PhysRevC.42.1838}, $N-\Omega$~\cite{IRITANI2019284}, $\Omega-\Omega$~\cite{PhysRevLett.120.212001}, respectively,
\begin{eqnarray}
\begin{split}
V_{np}(r) &= \sum_{i=1}^{2}C_i\frac{e^{-\mu_ir}}{r},\\
V_{N\Omega}(r) &= b_1e^{-b_2r^2} + b_3\left(1-e^{-b_4r^2}\right)\left(\frac{e^{-m_{\pi}r}}{r}\right)^2,\\
V_{\Omega\Omega}(r) &= \sum_{i=1}^{3} C_ie^{-(r/d_i)^2}.
\end{split}
\label{potential_def}
\end{eqnarray} 
Here  the parameters for the potentials are listed in Table~\ref{tab:potentialPar}.

\begin{table}
\caption{ \label{tab:potentialPar} Parameters for the potentials of $V_{np}(r)$~\cite{PhysRevC.42.1838}, $V_{N\Omega}(r)$~\cite{IRITANI2019284} and $V_{\Omega\Omega}(r)$~\cite{PhysRevLett.120.212001}. Note that the mass of pion is set to $m_{\pi}$ = 146 MeV.}
\begin{ruledtabular}
\begin{tabular}{lllll}

$V_{np}(r)$  & $C_1~(\text{MeV})$ & $C_2~(\text{MeV})$ & $\mu_1~(fm^{-1})$ & $\mu_2~(fm^{-1})$\\
&-626.885 & 1438.72 & 1.55 & 3.11\\
\hline
$V_{N\Omega}(r)$ & $b_1$ (MeV) & $b_2~(fm^{-2})$ & $b_3~(\text{MeV}\cdot fm^2)$ & $b_4~(fm^{-2})$\\
& -313 & 81.7 & -252 & 0.85\\
\hline
$V_{\Omega\Omega}(r)$ & $C_1$ (MeV)& $C_2$ (MeV) & $C_3$ (MeV) \\
& 914 & 305 & -112\\
& $d_1~(fm)$ & $d_2~(fm)$ & $d_3~(fm)$\\
 & 0.143 & 0.305 & 0.949\\
 
\end{tabular} 
\end{ruledtabular}
\end{table}

\begin{figure}[htb]
\includegraphics[angle=0,scale=0.44]{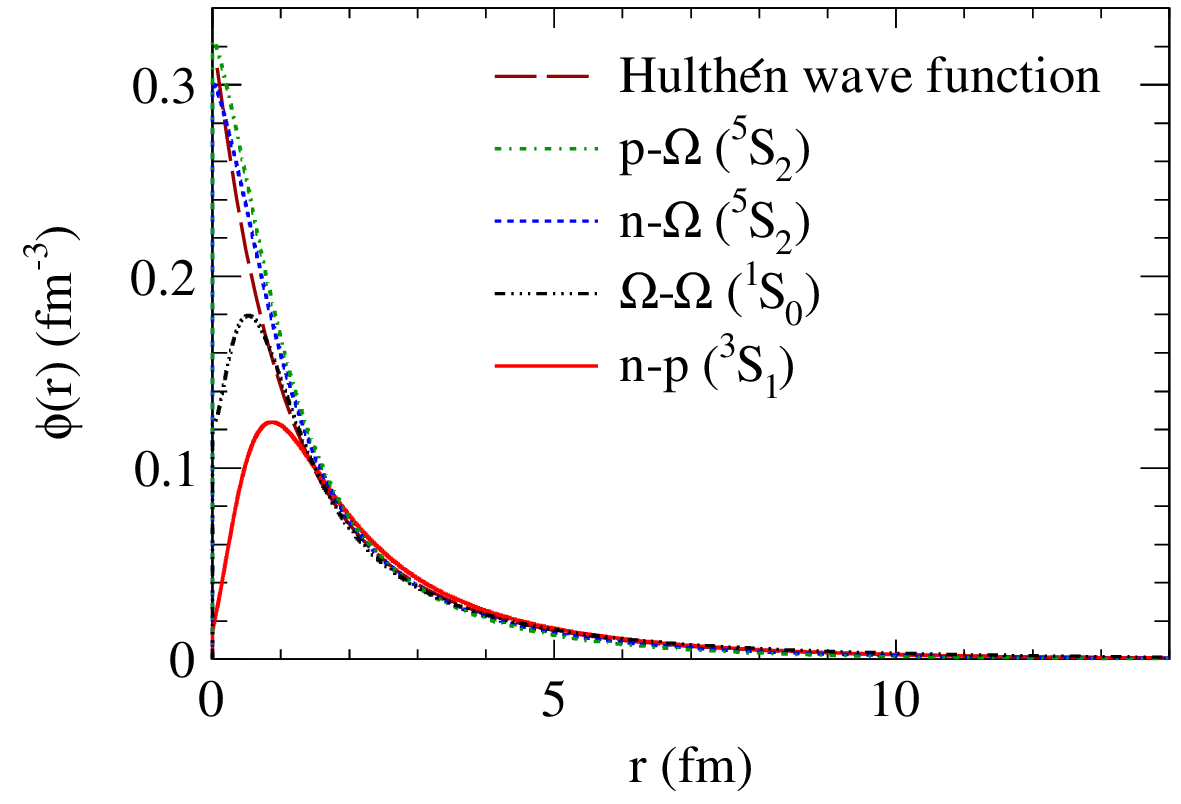}
\caption{The calculated relative wave functions of $d$ ($n$-$p$), $N\Omega$ and $\Omega\Omega$. The Hulth\'en wave function~\cite{CHEN2003809,nla.cat-vn2230793} for $d$ is also plotted.  For $p\Omega$ and $\Omega\Omega$, the Coulomb interaction was  taken into account.
\label{fig:WFun}(color online)
}
\end{figure}

In our calculation the Coulomb interaction was also taken into account for the charged pairs by adding $\pm\alpha/r$ ($+$ for $\Omega\Omega$ and $-$ for $p\Omega$) with $\alpha$ = $e^2/4\pi$ to the potential in Eq.~(\ref{potential_def}). Figure~\ref{fig:WFun} shows the numerical results of the wave function $\phi(r)$ for $d$ ($^3S_1$), $N\Omega$ ($^5S_2$) and $\Omega\Omega$  ($^1S_0$). The Hulth\'en wave function~\cite{CHEN2003809,nla.cat-vn2230793} for $d$ which is also presented in Fig.~\ref{fig:WFun} is higher than the calculated wave function by using the potential $V_{np}(r)$ in the short relative distance region because of the repulsion core of the potential for $n-p$ interaction, and this pattern was also found in $\Omega\Omega$ wave function. The $N-\Omega$ attractive potential results  that the wave function was similar to the $d$'s Hulth\'en wave function. Table~\ref{tab:Eb} shows the calculated binding energies which were consistent with the collected published results~\cite{PhysRevC.42.1838,IRITANI2019284,PhysRevLett.120.212001}. As discussed in Ref.~\cite{IRITANI2019284}, the pion mass could result in  a little discrepancy of the binding energy.  In this calculation, the masses of pion, nucleon and $\Omega$ were set as $m_{\pi}$ = 146 MeV, $m_{N}$ = 938 MeV and $m_{\Omega}$ = 1672 MeV, respectively.

The Wigner density function can be obtained through the Winger transformation by Eq.~(\ref{wigner_density_def}) from the resolved relative wave functions. In practice, it is found that the Wigner density function only depends on the values of relative coordinate $r$, relative momentum $q$ and the angle $\theta_{rq}$ between $\vec{r}$ and $\vec{q}$ after integration of azimuthal angle is performed analytically for the asymmetry of the relative wave function. The relative coordinate and momentum $\vec{r}(r, \phi_r, \theta_r)$ and $\vec{q}(q, \phi_q, \theta_q)$ can be rotated to, in a new frame, $\vec{r}(r, 0, 0)$ and $\vec{q}(q, \phi_q, \theta_{rq})$ with $\cos(\theta_{rq}) = \sin(\theta_r)\cos(\phi_r-\phi_q) + \cos(\theta_r)\cos(\theta_q)$ to keep the angle $\theta_{rq}$ between $\vec{r}$ and $\vec{q}$ unchanged. Then $\phi_R$ of $\vec{R}(R, \phi_R, \theta_R)$ will only appear in $\exp(-i\vec{q}\cdot\vec{R})$ and the integration in Eq.~(\ref{wigner_density_def}) can be written as the following equation  after integration of azimuthal angle,
 \begin{widetext}
\begin{eqnarray}
\begin{split}
\rho^W_2(r,q,\theta_{rq}) =&  \int \phi\left(\sqrt{r^2+\frac{R^2}{4}+\frac{rR}{2}\cos(\theta_R)}\right)\phi^{*}\left(\sqrt{r^2+\frac{R^2}{4}-\frac{rR}{2}\cos(\theta_R)}\right)\\
&\times2\pi J_0(qR\sin(\theta_R)\sin(\theta_{rq}))\cos(qR\cos(\theta_R)\cos(\theta_{rq})) \times\sin(\theta_R)d\theta_R R^2dR,
\end{split}
\label{wigner_density_def_int_phi}
\end{eqnarray}
\end{widetext}
here, $J_0(x)$ is the Bessel function of the first kind for order 0.

\begin{table}
\caption{ \label{tab:Eb} The  calculated binding energy together with the collected values from other published results~\cite{PhysRevC.42.1838,IRITANI2019284,PhysRevLett.120.212001}. Note that the masses of nucleon and $\Omega$ are set to $m_{N}$ = 938 MeV and $m_{\Omega}$ = 1672 MeV, respectively.}
\begin{ruledtabular}
\begin{tabular}{lll}
& this work & value/Reference\\
\hline
$E_d$ (MeV) & 2.23 & 2.2307~\cite{PhysRevC.42.1838}\\
$E_{p\Omega}$ (MeV) & 2.26 & 2.46~\cite{IRITANI2019284}\\
$E_{n\Omega}$ (MeV) & 1.38 & 1.54~\cite{IRITANI2019284}\\
$E_{\Omega\Omega}$ (MeV) &0.6 & 0.7~\cite{PhysRevLett.120.212001}\\

\end{tabular} 
\end{ruledtabular}
\end{table}

\section{Results and discussion}

\subsection{Blast-wave model + coalescence model (BLWC)}

\begin{figure*}[htb]
\includegraphics[angle=0,scale=0.90]{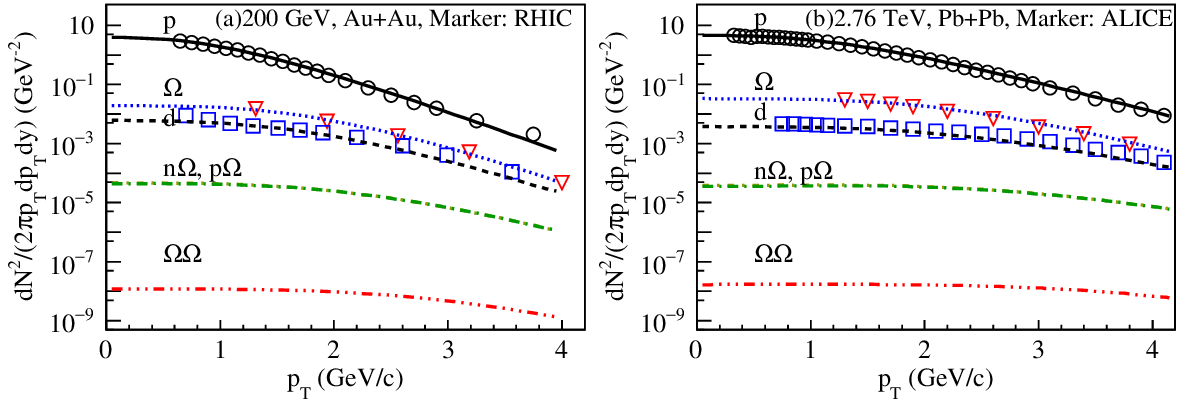}
\caption{Transverse momentum $p_T$ spectra of $p$, $d$, $\Omega$, $n\Omega$, $p\Omega$ and $\Omega\Omega$ in Au+Au collisions at $\sqrt{s_{NN}}$ = 200 GeV (a) and in Pb+Pb collisions at $\sqrt{s_{NN}}$ = 2.76 TeV  (b), respectively.  Lines for $p$ and $\Omega$: direct fits to the data; Lines  for $d$ and dibaryons ($p\Omega$, $n\Omega$ and $\Omega\Omega$): Blast-wave model coupled with dynamical coalescence model (BLWC); Markers: data from the RHIC~\cite{PhysRevC.69.034909,PhysRevC.99.064905,PhysRevLett.98.062301} and the ALICE~\cite{PhysRevC.88.044910,PhysRevC.93.024917,PLBAlice2014216}.
\label{fig:pTDst}(color online)
}
\end{figure*}

In Blast-wave model calculations~\cite{PhysRevC.70.044907,SUN2015272,PhysRevC.95.044905,PhysRevC.89.034918}, the parameters of $T_{kin}$, $\rho_0$, $\tau_0$, $\Delta\tau$ and $R_0$ can be obtained by fitting experimental transverse momentum $p_T$ spectra of $p$ and $\Omega$ by using Eq.~(\ref{pT_spectra_def}), and the parameters of $\tau_0$, $\Delta\tau$ and $R_0$ can be further  adjusted by fitting deuteron's $p_T$ spectra from experiments by using Eq.~(\ref{coal_ana_def}). In Au + Au collisions at $\sqrt{s_{NN}}$ = 200 GeV, the data were collected for $p$ from the PHENIX experiment~\cite{PhysRevC.69.034909}, for $\Omega$ and $d$ from the STAR experiments~\cite{PhysRevC.99.064905,PhysRevLett.98.062301} (centrality:  5\% for $p$, 0-10\% for $\Omega$ and $d$). In Pb+Pb collisions at $\sqrt{s_{NN}}$ = 2.76 TeV, the data of $p$, $\Omega$ and $d$ were taken from the ALICE experiments~\cite{PhysRevC.88.044910,PhysRevC.93.024917,PLBAlice2014216}  (centrality: 5\% for $p$, 0-10\% for $\Omega$ and $d$). From the fits to the experimental data of $p$, $\Omega$ and $d$ which are  shown in Fig.~\ref{fig:pTDst},  
the extracted parameters were 
$R_0$ = 12 $fm$, $\tau_0$ = 9 $fm/c$, $\Delta\tau$ = 3.5 $fm/c$, $T_{kin}$ = 111.6 MeV, and $\rho_0$ = 0.98 (0.9) for proton ($\Omega$) in Au+Au collisions at $\sqrt{s_{NN}}$ = 200 GeV,  
and 
$R_0$ = 19.7 $fm$, $\tau_0$ = 15.5 $fm/c$, $\Delta\tau$ = 1 $fm/c$, $T_{kin}$ = 122 MeV,  $\rho_0$ = 1.2 (1.07) for proton ($\Omega$) in Pb+Pb collisions at $\sqrt{s_{NN}}$ = 2.76 TeV. 

By using the above configured Blast-wave model and coalescence model (BLWC) as in Eq.~(\ref{coal_ana_def}), the transverse momentum $p_T$ spectra of $n\Omega$, $p\Omega$ and $\Omega\Omega$ dibaryons were calculated and shown in Fig.~\ref{fig:pTDst}(a) for  Au+Au collisions at $\sqrt{s_{NN}}$ = 200 GeV and Fig.~\ref{fig:pTDst}(b) for Pb+Pb collisions at $\sqrt{s_{NN}}$ = 2.76 TeV. The $p_T$ integrated yields $dN/dy$ of objects at midrapidity were given in Table~\ref{tab:dNdy_midrap} and the calculated $dN/dy$ of $p$, $\Omega$ and $d$ were comparable with those from experimental results from the RHIC data ~\cite{PhysRevC.69.034909,PhysRevC.99.064905,PhysRevLett.98.062301} as well as from the ALICE data  ~\cite{PhysRevC.88.044910,PhysRevC.93.024917,PLBAlice2014216}. The predicted $dN/dy$ of $p\Omega$, $n\Omega$ and $\Omega\Omega$ were $7.51\times 10^{-4}$, $7.39\times 10^{-4}$ and $0.31\times 10^{-6}$, respectively, for Au + Au collisions at the RHIC top energy,  and $1.31\times 10^{-3}$, $1.27\times 10^{-3}$ and $0.79\times 10^{-6}$ for Pb + Pb collisions at the ALICE  energy of 2.76 TeV, respectively. It is seen that the productions of $N\Omega$ and $\Omega\Omega$ at the ALICE energy were about 2 times of those at the RHIC top energy. These calculated results were similar to the previous work by using the naive coalescence model~\cite{SHAH20166} or analytical coalescence model~\cite{PhysRevC.95.044905} as well as using the AMPT model with $\Omega\Omega$ production channel~\cite{PAL2005210}.  

\begin{table}
\caption{ \label{tab:dNdy_midrap} Values of $dN/dy$ for $p\Omega$, $n\Omega$ and $\Omega\Omega$ dibaryons at mid-rapidity.}
\begin{ruledtabular}

\begin{tabular}{lllllll}
 & $p\Omega$ & $n\Omega$ & $\Omega\Omega$\\
 200 GeV\\
BLWC & 7.51$\times10^{-4}$ & 7.39$\times10^{-4}$ & 0.31$\times10^{-6}$\\
AMPTC & 9.5$\times10^{-4}$ & 9.5$\times10^{-4}$ & 0.81$\times10^{-6}$\\
 & $\pm 7.9\times10^{-5}$ & $\pm 7.9\times10^{-5}$ & $\pm1.1\times10^{-6}$\\
\hline
2.76 TeV&\\
BLWC & 1.31$\times10^{-3}$ & 1.27$\times10^{-3}$ & 0.79$\times10^{-6}$\\
AMPTC & 1.11$\times10^{-3}$ & 1.10$\times10^{-3}$ & 1.1$\times10^{-6}$\\
& $\pm8.57\times10^{-5}$ & $\pm8.57\times10^{-5}$ & $\pm1.3\times10^{-6}$\\
\end{tabular} 

\end{ruledtabular}
\end{table}

\begin{figure*}[htb]
\includegraphics[angle=0,scale=0.90]{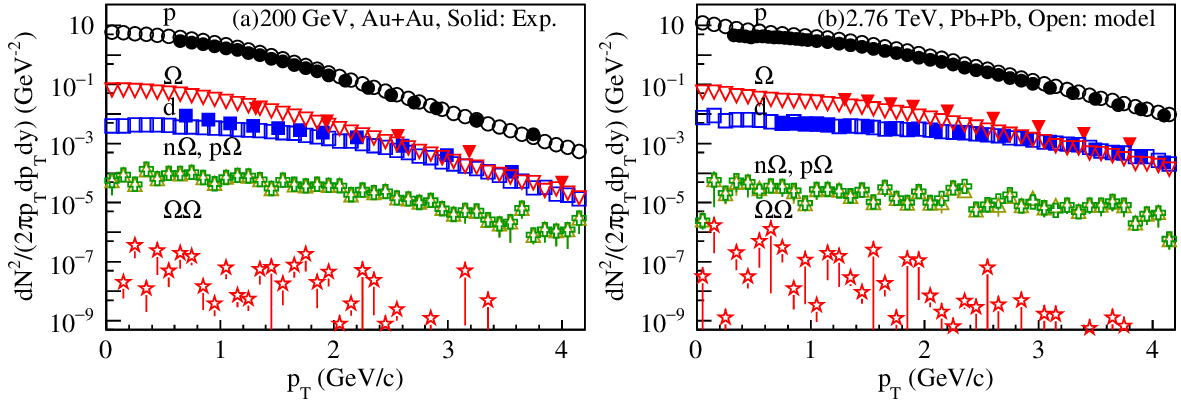}
\caption{Transverse momentum $p_T$ spectra of $p$, $d$, $\Omega$, $n\Omega$, $p\Omega$ and $\Omega\Omega$ in Au+Au collisions at $\sqrt{s_{NN}}$ = 200 GeV  (a) and in Pb+Pb collisions at $\sqrt{s_{NN}}$ = 2.76 TeV (b), respectively. 
Open markers for $p$ and $\Omega$: direct fits to the data; 
Open markers for $d$ and dibaryons ($p\Omega$, $n\Omega$ and $\Omega\Omega$): AMPT coupled with dynamical coalescence model (AMPTC); 
Solid markers: data from the RHIC~\cite{PhysRevC.69.034909,PhysRevC.99.064905,PhysRevLett.98.062301} and the ALICE~\cite{PhysRevC.88.044910,PhysRevC.93.024917,PLBAlice2014216}.
\label{fig:pTDstAMPT}(color online)
}
\end{figure*}

\subsection{AMPT model + coalescence model (AMPTC)}

The productions of $p\Omega$, $n\Omega$ and $\Omega\Omega$ bound states were also calculated by using phase-space data from the AMPT model~\cite{AMPT2005} via dynamical coalescence mechanism ~(\ref{coal_tranport_def}) (AMPTC).  The AMPT model gave the kinetic freeze-out position and momentum of each of particles ($p$, $n$ and $\Omega$ used here) at their freeze-out times and by using Eq.~(\ref{coal_tranport_def}) to coalesce into a dibaryon whose  relative coordinate and momentum were obtained after free streaming the first freeze-out constituent to the later one as did in Ref.~\cite{2020arXiv200500182S}, which was important for $N\Omega$ because the $\Omega$ was always freezed-out earlier than nucleons.  To fit proton spectra, some parameters defined in the original AMPT model~\cite{AMPT2005,AMPTGLM2016} were adjusted as $(a,b)$ = $(0.55, 0.1)$ for the RHIC energy and $(0.21, 0.075)$ for the LHC energy (here $a$ and $b$ are the Lund string fragmentation parameters defined in Ref.~\cite{AMPT2005}).  The coalescence mechanism for $\Omega$ was also developed as in Ref.~\cite{PhysRevC.100.064909} to fit $\Omega$ spectra. Fig.~\ref{fig:pTDstAMPT} presents the fitted $p_T$ spectra for proton and $\Omega$  as well as the coalesced $p_T$ spectra of $p\Omega$, $n\Omega$ and $\Omega\Omega$ bound states in Au + Au central collisions at $\sqrt{s_{NN}}$ = 200 GeV (a) and in Pb+Pb central collisions at $\sqrt{s_{NN}}$ = 2.76 TeV (b), respectively. Based on the adjusted AMPT parameters and the developed coalescence mechanism for $\Omega$, the $p_T$ spectra of $p$ and $\Omega$ could be described well and the $p_T$ spectra of $p\Omega$, $n\Omega$ and $\Omega\Omega$ were similar to those via the above BLWC calculation  shown in Fig.~\ref{fig:pTDst}. Values of $dN/dy$ for  $p\Omega$, $n\Omega$ and $\Omega\Omega$ were also listed in Table~\ref{tab:dNdy_midrap} and are consistent with those from the BLWC results.

In this work the particle interaction potential from the lattice QCD near physical pion mass $m_\pi$ = 146 MeV was taken into account and the relative wave function was assumed to be the $S-$wave function calculated by solving Schr\"odinger equation. By using the phase-space information from either the Blast-wave model or  the AMPT  model, the coalesced $p\Omega$, $n\Omega$, $\Omega\Omega$ gave the similar results and were in a good agreement with the previous prediction~\cite{SHAH20166,PhysRevC.95.044905,PAL2005210}. These consistent results implied that $N\Omega$ and $\Omega\Omega$ could be bounded in the $S-$wave state and produced via coalescence mechanism at final stage in relativistic heavy ion collisions. It also indicates that the hyperon-hyperon ($YY$) and hyperon-nucleon ($YN$) interactions from the first principle calculation, such as the lattice QCD, could be examined by investigating the production of $\Omega$-dibaryons as well as  $\Lambda$-dibaryons etc. 

\section{Summary}
The relative $S-$wave function was calculated through solving Schr\"odinger equation with the $N-\Omega$ and $\Omega-\Omega$ potential from the lattice QCD near physical pion mass $m_\pi$ = 146 MeV which was recently published by the HAL QCD collaboration, the calculated binding energy is consistent with the published results. In the coalescence mechanism frame, the Blast-wave model and AMPT model are respectively employed to provide the phase-space information of $p$ and $\Omega$ to coalesce into $N\Omega$ and $\Omega\Omega$ bound state by using the Wigner density function from the calculated relative $S-$wave function, and the production rate of the $\Omega$-dibaryons bound states agrees with other model prediction results. Although the Blast-wave model is very different from the AMPT model, these two models coupling with dynamical coalescence model give the similar results of these dibaryons. The present dynamical coalescence calculations of the $\Omega$-dibaryons shed light on the experimental searching for the (most)-strangeness dibaryon bound states at the STAR and the ALICE experiments, which  helps us to understand the $YY$ and $YN$ interactions.

{\it Acknowledgments}: This work was supported in part by the National Natural Science Foundation of China under contract Nos. 11421505, 11875066, 11890714, 11925502 and 11961141003, National Key R\&D Program of China under Grant No. 2016YFE0100900 and 2018YFE0104600, the Key Research Program of Frontier Sciences of the CAS under Grant No. QYZDJ-SSW-SLH002, and the Key Research Program of the CAS under Grant NO. XDPB09.



\bibliography{mybibfile}
\end{CJK*}	
\end{document}